\documentclass [12pt]{article}
\pdfoutput=1

\usepackage{amsmath}
\usepackage{amsfonts}
\usepackage{amscd}
\usepackage{amsthm}
\usepackage{setspace}

\usepackage{graphicx}
\usepackage{authblk}
\usepackage{caption}
\usepackage{ytableau}
\usepackage{mathtools}

\def\Z{\mathbb{Z}}

\def\P{\mathbb{P}}

\begin{document}

\begin{titlepage}

\begin{flushright}
KEK-TH-2218
\end{flushright}

\vskip 3.0cm

\begin{center}

{\bf \Large Singularities of 1/2 Calabi--Yau 4-folds and classification scheme for gauge groups in four-dimensional F-theory}

\vskip 1.2cm

Yusuke Kimura$^1$ 
\vskip 0.6cm
{\it $^1$KEK Theory Center, Institute of Particle and Nuclear Studies, KEK, \\ 1-1 Oho, Tsukuba, Ibaraki 305-0801, Japan}
\vskip 0.4cm
E-mail: kimurayu@post.kek.jp

\vskip 2cm
\abstract{In a previous study, we constructed a family of elliptic Calabi--Yau 4-folds possessing a geometric structure that allowed them to be split into a pair of rational elliptic 4-folds. In the present study, we introduce a method of classifying the singularity types of this class of elliptic Calabi--Yau 4-folds. In brief, we propose a method to classify the non-Abelian gauge groups formed in four-dimensional (4D) $N=1$ F-theory for this class of elliptic Calabi--Yau 4-folds. 
\par To demonstrate our method, we explicitly construct several elliptic Calabi--Yau 4-folds belonging to this class and study the 4D F-theory thereupon. These constructions include a 4D model with two U(1) factors.} 

\end{center}
\end{titlepage}

\tableofcontents
\section{Introduction}
U(1) gauge symmetry has been a subject of intensive study in F-theory. Compactifications spaces used in the formulation of F-theory \cite{Vaf, MV1, MV2} admit a genus-one fibration, enabling the axio-dilaton to exhibit an $SL(2, \Z)$ monodromy. When a genus-one fibration possesses a global section \footnote{F-theory models of elliptic fibrations possessing a global section have been analyzed. Recent studies of such models can be found, e.g., in \cite{MorrisonPark, MPW, BGK, BMPWsection, CKP, BGK1306, CGKP, CKP1307, CKPS, HLY2013, Mizoguchi1403, AL, EKY1410, LSW, CKPT, CGKPS, HJP2015, MP2, MPT1610, BMW2017, CL2017, KimuraMizoguchi, Kimura1802, AHMT1803, LRW2018, TasilectWeigand, MizTani2018, TasilectCL, Kimura1810, CMPV1811, H2018, TT2019, Kimura1902, Kimura1903, EJ1905, LW1905, Kimura1910, CKS1910, Kimura1911, FKKMT1912, AFHRZ2001, Kimura2003, KMT2003, Kimura2003subextremal, HHP2020, DHRT2020}. F-theory model constructions where one or more U(1) factors are formed are discussed, for example, in \cite{MorrisonPark, BMPWsection, CKP, CGKP, BMPW2013, CKPS, MTsection, MT2014, KMOPR, BGKintfiber, CKPT, GKK, MPT1610, LS2017, Kimura1802, TT2018, CLLO, CMPV1811, TT2019, Kimura1908, Kimura1910, Kimura1911, OS2019, AFHRZ2001, Kimura2003subextremal}.}, the set of global sections that the genus-one fibration admits form a group, known in mathematics as the ``Mordell--Weil group.'' The rank of the Mordell--Weil group is known to yield the number of U(1) gauge group factors formed in F-theory when compactified on that elliptic fibration \cite{MV2}. 
\par A family of elliptically fibered Calabi--Yau 4-folds, possessing a structure such that they can be split into a pair of rational elliptic 4-fold building blocks, was constructed in \cite{Kimura1911}. The rational elliptic 4-fold building blocks of such a family of Calabi--Yau 4-folds are referred to as ``1/2 Calabi--Yau 4-folds'' \cite{Kimura1911}. One motivation for introducing 1/2 Calabi--Yau 4-folds in \cite{Kimura1911} was that various numbers of U(1) gauge group factors \footnote{See, e.g., \cite{KCY4, Kdisc} for explicit constructions of four-dimensional $N=1$ F-theory models without a U(1) gauge group.} are formed in four-dimensional (4D) F-theory on the elliptically fibered Calabi--Yau 4-folds built as double covers of the 1/2 Calabi--Yau 4-folds. The general structures of the 1/2 Calabi--Yau 4-folds and the Calabi--Yau 4-folds built as their double covers, as well as the numbers of U(1) factors formed in 4D F-theory on the Calabi--Yau 4-fold double covers, have been analyzed in \cite{Kimura1911}. However, analyses of the non-Abelian gauge groups and matter spectra formed in 4D F-theory, as well as the classification of 1/2 Calabi--Yau 4-folds, were left for future studies. 
\par Explicit constructions of 1/2 Calabi--Yau 4-folds with no $ADE$ singularity type were given in \cite{Kimura1911}. At least six U(1) factors are formed in 4D $N=1$ F-theory on the Calabi--Yau 4-folds built as the double covers of 1/2 Calabi--Yau 4-folds with no $ADE$ singularity, and the resulting theories do not possess a non-Abelian gauge group factor \cite{Kimura1911}. 
\par There are two motivations for this study: \\
i) Constructions of F-theory models discussed in \cite{Kimura1911} provide a series of 4D theories with various numbers of U(1) gauge group factors. However, as we mentioned previously an approach to classify the non-Abelian gauge groups formed in the theories was not discussed in \cite{Kimura1911}. In this study, we would like to provide a method to classify them. \\
ii) We aim to introduce a technique to determine the structure of a 1/2 Calabi--Yau 4-fold, when its singularity type is given. Concretely, four (1,1) hypersurfaces in $\P^2\times\P^2$ control the structure of a 1/2 Calabi--Yau 4-fold. We provide a method to deduce the equations of the four hypersurfaces from the singularity type of a 1/2 Calabi--Yau 4-fold \footnote{This also works in the reversed direction. Namely, our method can also be applied to deduce the singularity type of a 1/2 Calabi--Yau 4-fold when the equations of four (1,1) hypersurfaces are given.}. This also determines, in principle, the structure of the Calabi--Yau 4-fold double cover. This approach can aid in deducing the non-Abelian gauge groups formed in 4D F-theory, as well as the matter spectra localized at the intersections of the 7-branes, and Yukawa couplings.
\par Here, we develop a method of extracting information of the non-Abelian gauge groups formed in 4D F-theory on the Calabi--Yau 4-fold double covers of 1/2 Calabi--Yau 4-folds. In this paper, we provide a classification scheme for the singularity types of 1/2 Calabi--Yau 4-folds. We also provide some explicit constructions of 1/2 Calabi--Yau 4-folds with $ADE$ singularity types. A 1/2 Calabi--Yau 4-fold and the Calabi--Yau 4-fold constructed as its double cover possess identical singularity types \cite{Kimura1911}; this yields a method of classifying the singularity types of the Calabi--Yau 4-folds built as double covers of the 1/2 Calabi--Yau 4-folds. In the language of string theory, this provides a method of classifying the types of non-Abelian gauge groups formed on the 7-branes \cite{MV2, BIKMSV} in 4D F-theory on the Calabi--Yau 4-folds constructed as double covers of the 1/2 Calabi--Yau 4-folds. The number of U(1) factors formed in 4D F-theory can be deduced from the ranks of the singularity types, using the method discussed in \cite{Kimura1911}. 
\par To classify the singularity types of the 1/2 Calabi--Yau 4-folds (and those of their Calabi--Yau 4-fold double covers), we apply the techniques discussed in the elegant and interesting work of Shigeru Mukai \cite{Muk, Mukai2008, Mukai2019}. With these techniques, the classification of the singularity types of the 1/2 Calabi--Yau 4-folds reduces to that of cubic hypersurfaces in $\P^3$. Furthermore, we use blow-ups to analyze the structures of the singular fibers \footnote{We utilize Kodaira's notation \cite{Kod1, Kod2} to denote the types of the singular fibers. The classification of the types of the singular fibers of elliptically fibered surfaces can be found in \cite{Kod1, Kod2}, and techniques that determine the fiber types of elliptically fibered surfaces are discussed in \cite{Ner, Tate}.} corresponding to the singularity types. These can be used to study the matter fields localized at the intersections of the 7-branes, as well as the Yukawa couplings. 
\par The ranks of the singularity types of the 1/2 Calabi--Yau 4-folds vary from zero to six. The singularity rank and Mordell--Weil rank of any 1/2 Calabi--Yau 4-fold sum to six. The Mordell--Weil rank of a Calabi--Yau 4-fold double cover is greater than or equal to the Mordell--Weil rank of the original 1/2 Calabi--Yau 4-fold. These properties have been proved in \cite{Kimura1911}. Owing to these properties of the 1/2 Calabi--Yau 4-folds, the number of U(1) gauge group factors formed in 4D F-theory on the Calabi--Yau 4-fold double covers can be deduced \cite{Kimura1911}. While it was demonstrated in \cite{Kimura1911} that the singularity types of an original 1/2 Calabi--Yau 4-fold and its Calabi--Yau 4-fold double cover are identical, the $ADE$ classification of the singularity types was not given in \cite{Kimura1911}; in this work, we demonstrate that applying the techniques presented in \cite{Muk, Mukai2008, Mukai2019} yields a method to classify them, and that the types of the non-Abelian gauge groups formed in 4D F-theory can be deduced. 
\par Our classification scheme can be used to study the gauge groups in a series of 4D $N=1$ F-theory models, in which various numbers of U(1) factors are formed. 
\par To explicitly demonstrate our method, we construct Calabi--Yau 4-fold double covers of 1/2 Calabi--Yau 4-folds, possessing $3A_2$ and $D_4$ singularities. 
\par We identify the ``puzzling problem'' of whether a 1/2 Calabi--Yau 4-fold with an $E_6$ singularity exists or not. While this does not suggest a mathematical inconsistency, the existence is left undetermined owing to technical issues. Consequently, the question of whether an elliptic Calabi--Yau 4-fold with an $E_6$ singularity (fibered over a Fano 3-fold of degree-two) exists is left open. We discuss this ``puzzle'' at length in section \ref{sec2.4}. 

\vspace{5mm}

\par Local models of F-theory model constructions \cite{DWmodel, BHV1, BHV2, DW} have been emphasized in recent studies. However, the global aspects of the compactification geometry need to be analyzed before discussing the issues pertaining to gravity and the early universe. In this work, we study the geometry of Calabi--Yau 4-folds from the global perspective.
\par The four-form flux \cite{BB, SVW, W96, GVW, DRS} contributes several effects that can alter the gauge groups and matter spectra in 4D F-theory \cite{MTsection}. However, when four-form flux is turned on in our 4D F-theory constructions, the cohomology groups of the Calabi--Yau 4-folds built as the double covers of the 1/2 Calabi--Yau 4-folds need to be analyzed to study the effects of the four-form flux. In this study, we do not discuss the situation in which the four-form flux is turned on.
\par One of the problems raised in section \ref{sec3} can be possibly related to the swampland conditions. The problem possibly related to the swampland conditions mentioned in section \ref{sec3} is concerning the number of gauge group factors formed in 4D F-theory constructions \footnote{The numbers of gauge group factors formed in 4D F-theory on elliptic Calabi--Yau 4-folds over toric 3-folds were discussed in \cite{TW2015}}. Reviews of recent studies on the swampland conditions can be found in \cite{BCV1711, Palti1903}. The notion of the swampland was discussed in \cite{Vafa05, AMNV06, OV06}. 

\vspace{5mm}

\par This paper is structured as follows. A summary of the results obtained in the study is provided in section \ref{sec2.1}. Our strategy for classifying the singularity types of the 1/2 Calabi--Yau 4-folds and their Calabi--Yau 4-fold double covers are also discussed. Explicit constructions of 1/2 Calabi--Yau 4-folds with rank-six and rank-four singularities are given in sections \ref{sec2.2} and \ref{sec2.3}, respectively. The existence of the 1/2 Calabi--Yau 4-fold with an $E_6$ singularity is left undetermined owing to some technical issues; this case is discussed in section \ref{sec2.4}. 4D F-theory models on the Calabi--Yau 4-folds constructed as double covers of the 1/2 Calabi--Yau 4-folds with singularities are studied in section \ref{sec3}. In section \ref{sec4}, we state our concluding remarks and highlight the problems that remain unresolved.

\section{Classification scheme}
\label{sec2}
\subsection{A general strategy}
\label{sec2.1}
1/2 Calabi--Yau 4-folds are constructed as blow-ups of the product of the complex projective planes $\P^2\times \P^2$ at the six intersection points of four (1,1) hypersurfaces: $H_1, H_2, H_3, H_4$ \cite{Kimura1911}. (The intersection points are counted with multiplicity; the actual number of intersection points can be less than six.) Taking double covers of the 1/2 Calabi--Yau 4-folds, ramified along a degree-six polynomial in the variables of $H_1, H_2, H_3, H_4$, yields elliptically fibered Calabi--Yau 4-folds \cite{Kimura1911}. 
\par The general construction of the 1/2 Calabi--Yau 4-folds, as well as some of their characteristic properties (such as that the rank of the singularity type and the Mordell--Weil rank always sum to six), and those of the Calabi--Yau 4-folds constructed as their double covers, were studied in \cite{Kimura1911}; however, the classification of $ADE$ singularities of the 1/2 Calabi--Yau 4-folds was not discussed in \cite{Kimura1911}. The singularity types of the 1/2 Calabi--Yau 4-folds and their double covers are necessary for deducing the non-Abelian gauge groups formed in 4D $N=1$ F-theory on the resulting Calabi--Yau 4-folds.
\par We propose a method to classify the singularity types, employing the techniques described in \cite{Muk, Mukai2008, Mukai2019}. Because the singularity types of an original 1/2 Calabi--Yau 4-fold and its Calabi--Yau 4-fold double cover are identical \cite{Kimura1911}, it is only necessary to classify the singularity types of the 1/2 Calabi--Yau 4-folds. Several pairs of ``projective dual'' del Pezzo manifolds were studied in \cite{Mukai2019}. We used one such pair, $(X_3, Y_6)$. Here, $X_3$ denotes a cubic hypersurface in $\P^8$, and $Y_6$ is a Segre variety $\P^2\times \P^2$ embedded inside $\P^8$. We can consider four hypersurfaces, $F_1, F_2, F_3,$ and $F_4$, in $\P^8$ in such a way that when they are restricted to $Y_6$, they yield (1,1) hypersurfaces in $\P^2\times \P^2$. The blow-up of $\P^2\times\P^2$ at the intersection points of the four restricted hypersurfaces yields a 1/2 Calabi--Yau 4-fold, as previously mentioned. On the dual $X_3$ side, the projective duals of the four hypersurfaces yield four points in $\P^8$, spanning a $\P^3$ \cite{Mukai2019}. Therefore, on the $X_3$ side, the four hypersurfaces correspond to a cubic hypersurface in $\P^3$. 
\par This means that the configuration of the intersection points of four (1,1) hypersurfaces in $\P^2\times\P^2$, which yields a 1/2 Calabi--Yau 4-fold when the intersection points are blown up, corresponds to a cubic hypersurface in $\P^3$ via ``projective duality.''
\par Based on an argument similar to that given in \cite{Kimura2003}, and by utilizing properties of projective duality and the techniques described in \cite{Muk, Mukai2008, Mukai2019}, we conclude that the singularity types of the 1/2 Calabi--Yau 4-folds are identical to those of the cubic hypersurfaces in $\P^3$. 
\par The classification of the singularity types of cubic hypersurfaces in $\P^3$ can be found in \cite{DolgachevAlgGeom}. Based on this classification, in principle, we can classify the singularity types of the 1/2 Calabi--Yau 4-folds. 
\par However, there is a subtlety in this ``equivalence'' of singularity types. Given a cubic hypersurface in $\P^3$, we need a matrix representation of the cubic hypersurface to ensure that there exists a corresponding 1/2 Calabi--Yau 4-fold with an identical singularity type. Thus, the question of whether a 1/2 Calabi--Yau 4-fold with an $E_6$ exists or not is left undetermined. We discuss this problem in section \ref{sec2.4}.
\par When the matrix representation of a given cubic hypersurface in $\P^3$ can be determined, the equations of the four (1,1) hypersurfaces, $H_1, H_2, H_3,$ and $H_4$, in $\P^2\times\P^2$ that yield a 1/2 Calabi--Yau 4-fold corresponding to the cubic hypersurface can be deduced from the matrix representation. 
\par The structures of the singular fibers of the 1/2 Calabi--Yau 4-folds can be analyzed from the deduced equations of the four (1,1) hypersurfaces. In the analysis, a blow-up (or even multiple stages of them) need to be performed. Because the singularity types of an original 1/2 Calabi--Yau 4-fold and the Calabi--Yau 4-fold constructed as its double cover are identical \cite{Kimura1911}, the structures of the singular fibers of the Calabi--Yau 4-fold double cover can also be deduced in this manner. As a result, we obtain the types of the non-Abelian gauge groups formed in 4D F-theory on the Calabi--Yau 4-fold double covers. The analysis of the singular fibers described here may be used to study matter spectra at the intersections of the 7-branes. 
\par To demonstrate our method, we explicitly study the 1/2 Calabi--Yau 4-folds with rank-six \footnote{The singularity types of rank six are the highest of the 1/2 Calabi--Yau 4-folds \cite{Kimura1911}.} and rank-four singularities in sections \ref{sec2.2} and \ref{sec2.3}, respectively. 
\par According to the classification results in \cite{DolgachevAlgGeom}, there are three rank-six singularity types for cubic hypersurfaces in $\P^3$: $3A_2$, $A_5A_1$, and $E_6$. A 1/2 Calabi--Yau 4-fold possessing the first singularity type is constructed in section \ref{sec2.2} \footnote{1/2 Calabi--Yau 4-fold with an $A_5A_1$ singularity type can be constructed in a similar manner using our method.}. We identify a ``puzzle'' concerning whether 1/2 Calabi--Yau 4-folds with an $E_6$ singularity type exist or not, and this ``puzzle'' is discussed in section \ref{sec2.4}. 
\par We also explicitly construct a 1/2 Calabi--Yau 4-fold with a $D_4$ singularity type in section \ref{sec2.3}.
\par The Calabi--Yau 4-folds constructed as double covers of these 1/2 Calabi--Yau 4-folds, as well as the 4D $N=1$ F-theory on the Calabi--Yau 4-fold double covers, are studied in section \ref{sec3}.

\subsection{$3A_2$ singularity}
\label{sec2.2}
\begin{figure}
\begin{center}
\includegraphics[height=10cm, bb=0 0 960 540]{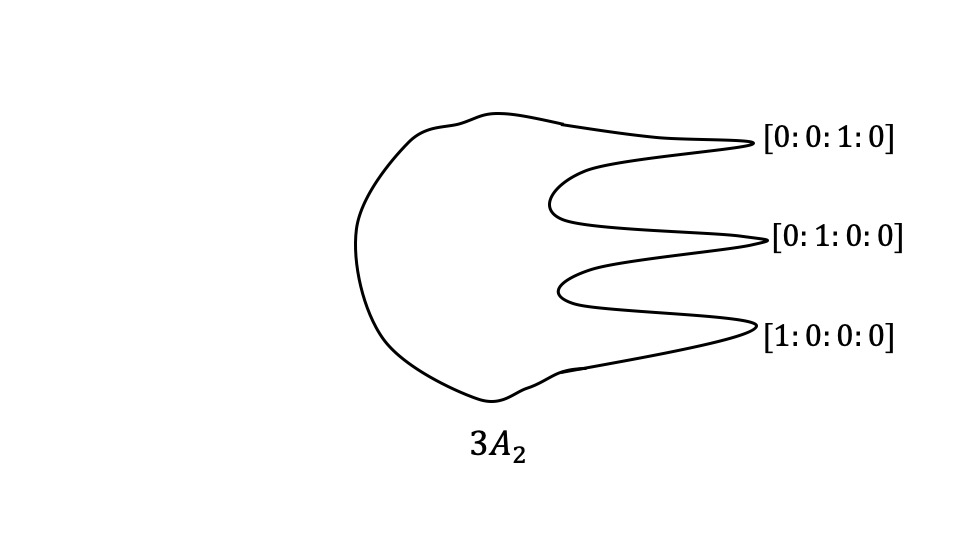}
\caption{\label{figure3A2}Image of a cubic hypersurface in $\P^3$ with three $A_2$ singularities.}
\end{center}
\end{figure}

We construct 1/2 Calabi--Yau 4-fold with a $3A_2$ singularity to demonstrate our method. After some consideration, we find that a dual cubic hypersurface with $3A_2$ singularity in $\P^3$ is given by the following equation:
\begin{equation}
\label{cubic hypersurface 3A2 in 2.2}
x_4^3+x_1x_2x_3=0,
\end{equation}
where we used $[x_1:x_2:x_3:x_4]$ to denote the homogeneous coordinates of $\P^3$. It can be explicitly seen that the cubic hypersurface (\ref{cubic hypersurface 3A2 in 2.2}) actually possesses three $A_2$ singularities. The derivation is as follows. We set $x_3=1$; then, the equation (\ref{cubic hypersurface 3A2 in 2.2}) becomes $x_4^3+x_1x_2=0$. This hypersurface has an $A_2$ singularity at the origin $(x_1,x_2,x_4)=(0,0,0)$. In a similar way, the other two $A_2$ singularities can be found by setting $x_1=1$ and $x_2=1$. An image of the cubic hypersurface with a $3A_2$ singularity is given in Figure \ref{figure3A2}. 

\par Then, we proceed to the determinantal representation of the cubic hypersurface (\ref{cubic hypersurface 3A2 in 2.2}); this is given by a 3 $\times$ 3 matrix \footnote{The matrix representation of a cubic hypersurface in $\P^3$ need not be symmetric, whereas the matrix representation of a quartic plane curve as the dual of a 1/2 Calabi--Yau 3-fold \cite{Kimura2003, Kimura2003subextremal} must be symmetric.} as follows:
\begin{equation}
\label{matrix rep 3A2 in 2.2}
\begin{pmatrix}
x_4 & x_1 & 0 \\
0 & x_4 & x_2 \\
x_3 & 0 & x_4 \\
\end{pmatrix}.
\end{equation}
The equations of the four (1,1) hypersurfaces yielding the 1/2 Calabi--Yau 4-fold with $3A_2$ singularity can be deduced from the matrix representation (\ref{matrix rep 3A2 in 2.2}). We use $[x:y:z]$ and $[s:t:u]$ to denote the homogeneous coordinates of the first and second $\P^2$s in $\P^2\times \P^2$, respectively. Entries in the matrix representation correspond to monomials of the (1,1) polynomial in $\P^2\times\P^2$. For example, the (2,2)th entry of the matrix representation corresponds to $yt$, and the (1,2)th entry corresponds to $xt$. The entries where $x_i$ appears correspond to the $i$th (1,1) hypersurface $H_i$, where $i=1,2,3,4$. Therefore, we find that the equations of the (1,1) hypersurfaces are given as follows:
\begin{eqnarray}
\label{four hypersurfaces 3A2 in 2.2}
H_1 = & xt \\ \nonumber
H_2 = & yu \\ \nonumber
H_3 = & zs \\ \nonumber
H_4 = & xs+yt+zu. 
\end{eqnarray}
A blow-up of $\P^2\times\P^2$ at the base points of the hypersurfaces (\ref{four hypersurfaces 3A2 in 2.2}) yields the 1/2 Calabi--Yau 4-fold with a $3A_2$ singularity. 
\par In section \ref{sec3}, we discuss the 4D F-theory on the Calabi--Yau 4-fold double covers of the 1/2 Calabi--Yau 4-folds considered in this section. 

\vspace{5mm}

\par Furthermore, we attempt to analyze the structure of the singular fibers by performing blow-ups; when the double cover of the 1/2 Calabi--Yau 4-fold is considered, this information is relevant to the non-Abelian gauge group factor formed in 4D F-theory and the matter fields localized at the intersections of the 7-branes. 
\par The four hypersurfaces (\ref{four hypersurfaces 3A2 in 2.2}) have three base points: $([1:0:0], [0:0:1])$, $([0:1:0], [1:0:0])$, and $([0:0:1], [0:1:0])$. 
\par The equation for the singular fibers of the resulting 1/2 Calabi--Yau 4-fold corresponding to the $A_2$ singularity of a dual cubic hypersurface (\ref{cubic hypersurface 3A2 in 2.2}) at the point $[0:0:1:0]$ is given as follows:
\begin{eqnarray}
\label{equation singular fiber in 2.2}
zs = & 0 \\ \nonumber
b\, xt-a\, yu = & 0 \\ \nonumber
c\, xt - a\, (xs+yt+zu) = & 0.
\end{eqnarray}
Here, $a,b,c$ denote the parameters of the discriminant component in the base of the 1/2 Calabi--Yau 4-fold over which the fibers become singular, corresponding to an $A_2$ singularity at $[0:0:1:0]$ in the cubic hypersurface. 
\par Because the first equation in (\ref{equation singular fiber in 2.2}) is reducible into two linear factors ($z$ and $s$), the equations in (\ref{equation singular fiber in 2.2}) describe two $\P^1$s meeting at two points. One of the two intersection points is one of the three base points of the four (1,1) hypersurfaces, $([1:0:0], [0:0:1])$. When this point is blown up, the two $\P^1$s are separated at the point $([1:0:0], [0:0:1])$, and $\P^1$ arises at this point as a result of the blow-up. The resulting structure of the singular fiber is given by three $\P^1$s, any pair of which meet at one point; that is, a type $I_3$ fiber. The situation is analogous to the appearance of the type $I_3$ fiber described in \cite{Kimura2003subextremal}. 
\par By a similar argument, it can be found that the other singular fibers corresponding to the remaining two $A_2$ singularities of the dual cubic hypersurface (\ref{cubic hypersurface 3A2 in 2.2}) are type $I_3$ fibers. 
\par One can also construct a cubic hypersurface with an $A_5A_1$ singularity in $\P^3$. By computing the matrix representation of this cubic hypersurface, the equations of the four (1,1) hypersurfaces can be deduced, yielding the dual 1/2 Calabi--Yau 4-fold with an $A_5A_1$ singularity type. 
\par Therefore, the existence of ``extremal'' 1/2 Calabi--Yau 4-folds with $3A_2$ and $A_5A_1$ singularity types \footnote{We refer to these 1/2 Calabi--Yau 4-folds as ``extremal'' 1/2 Calabi--Yau 4-folds, in the sense that they have the highest singularity rank.} can be confirmed constructively. However, the case of the $E_6$ singularity type is more complicated, as we discuss in section \ref{sec2.4}. 

\subsection{$D_4$ singularity}
\label{sec2.3}
A cubic hypersurface with a $D_4$ singularity in $\P^3$ is given by the following equation:
\begin{equation}
\label{hypersurface D4 in 2.3}
x_1^2x_3+x_3^3- x_2^2x_4=0.
\end{equation}
The $D_4$ singularity is located at $[0:0:0:1]$ \footnote{This can be found by setting $x_4=1$ and comparing the equation (\ref{hypersurface D4 in 2.3}) with the (local) equation of $D_4$ singularity given in, e.g., \cite{KM92}.}. By a standard argument it can be found that this is a unique singularity of the cubic hypersurface (\ref{hypersurface D4 in 2.3}). 
\par Through calculation, we learn that the determinantal representation of the cubic hypersurface with a $D_4$ singularity (\ref{hypersurface D4 in 2.3}) is given as follows:
\begin{equation}
\label{matrix rep D4 in 2.3}
\begin{pmatrix}
0 & -x_2 & x_3-i\, x_1 \\
x_2 & x_3 & 0 \\
x_3+i\, x_1 & 0 & x_4 \\
\end{pmatrix}.
\end{equation}
By using the determinantal representation (\ref{matrix rep D4 in 2.3}) and using a similar method to that employed in section \ref{sec2.2}, we can derive the equations of the four (1,1) hypersurfaces yielding the dual 1/2 Calabi--Yau 4-fold. We deduce that the equations of these four hypersurfaces are given as follows:
\begin{eqnarray}
H_1= & -i\, xu +i\, zs \\ \nonumber
H_2= & -xt+ ys \\ \nonumber
H_3 = & xu+yt+zs \\ \nonumber
H_4 = & zu.
\end{eqnarray}
\par The base points consist of three points: $([1:0:0], [1:0:0])$, $([0:1:0],[0:0:1])$, and $([0:0:1], [0:1:0])$. Blow-ups at the base points yield the 1/2 Calabi--Yau 4-fold with a $D_4$ singularity as the ``dual'' of the cubic hypersurface (\ref{hypersurface D4 in 2.3}). 
\par Because the resulting 1/2 Calabi--Yau 4-fold has a singularity rank of four, its Mordell--Weil rank is two.

\subsection{A puzzle}
\label{sec2.4}
As discussed in \cite{DolgachevAlgGeom}, the rank-six singularity types of cubic hypersurfaces in $\P^3$ are: $3A_2$, $A_5A_1$, and $E_6$. We constructed a 1/2 Calabi--Yau 4-fold corresponding to the first singularity type. The 1/2 Calabi--Yau 4-fold with an $A_5A_1$ singularity type can be constructed using a method similar to that given in section \ref{sec2.2}.  
\par There is a subtlety concerning the last singularity type, $E_6$. While a cubic hypersurface with an $E_6$ singularity in $\P^3$ exists, it does not allow for a matrix representation (Table 9.2 in \cite{DolgachevAlgGeom}). Therefore, our method does not (at least directly) apply to the $E_6$ singularity and does not determine whether a 1/2 Calabi--Yau 4-fold with an $E_6$ singularity exists or not. 
\par All other singularity types for cubic hypersurfaces in $\P^3$ have matrix representations \cite{DolgachevAlgGeom}; therefore, our method applies to the remaining singularity types, including the singularity types of ranks lower than six. The method shows that the corresponding 1/2 Calabi--Yau 4-folds do indeed exist. The $E_6$ singularity is an exception. If a 1/2 Calabi--Yau 4-fold with an $E_6$ singularity exists, a construction that does not rely on the matrix representation is needed. 
\par Is there a physical reason behind the $E_6$ singularity's ``special'' status? The existence of an elliptically fibered Calabi--Yau 4-fold with an $E_6$ singularity over a degree-two Fano 3-fold as base 3-fold is also undetermined. It may be interesting to study whether such 1/2 Calabi--Yau 4-folds and elliptic Calabi--Yau 4-folds over a degree-two Fano 3-fold possessing an $E_6$ singularity exist, and these topics are left for future studies.

\section{Applications to 4D F-theory}
\label{sec3}
Taking the double covers of 1/2 Calabi--Yau 4-folds, ramified over a degree-six polynomial in the variables of four (1,1) hypersurfaces, $H_1, H_2, H_3,$ and $H_4$, yields elliptically fibered Calabi--Yau 4-folds \cite{Kimura1911}. The base 3-folds of the resulting Calabi--Yau 4-folds are isomorphic to degree-two Fano 3-folds. F-theory compactifications on the Calabi--Yau 4-folds yield 4D $N=1$ theories. 
\par Because the singularity types of the Calabi--Yau 4-folds constructed as double covers are identical to those of the original 1/2 Calabi--Yau 4-folds, the types of non-Abelian gauge group factors formed in 4D F-theory can be derived from the singularity types of the 1/2 Calabi--Yau 4-folds. Particularly, F-theory on the Calabi--Yau 4-fold double covers of the 1/2 Calabi--Yau 4-folds constructed in sections \ref{sec2.2} and \ref{sec2.3} possess non-Abelian gauge group factors corresponding to $3A_2$ and $D_4$ singularity types. Owing to the equality that holds for 1/2 Calabi--Yau 4-folds (which states that the singularity rank and Mordell--Weil rank always sum to six \cite{Kimura1911}), the Mordell--Weil rank of the 1/2 Calabi--Yau 4-folds constructed in section \ref{sec2.2} is zero. Utilizing the property \cite{Kimura1911} that the Calabi--Yau 4-fold double cover of a 1/2 Calabi--Yau 4-fold has Mordell--Weil rank greater than or equal to that of the original 1/2 Calabi--Yau 4-fold does not provide any new information concerning the number of U(1) factors formed in 4D F-theory on the Calabi--Yau 4-fold as the double cover of the 1/2 Calabi--Yau 4-fold constructed in section \ref{sec2.2}. 
\par 1/2 Calabi--Yau 4-fold with a $D_4$ singularity (as constructed in section \ref{sec2.3}) has Mordell--Weil rank two; thus, the double cover of this 1/2 Calabi--Yau 4-fold yields an elliptic Calabi--Yau 4-fold with Mordell--Weil rank greater than or equal to two. Therefore, at least two U(1) factors are formed in the 4D F-theory on the Calabi--Yau 4-fold as the double cover of the 1/2 Calabi--Yau 4-fold constructed in section \ref{sec2.3}. 
\par Our method also applies to 1/2 Calabi--Yau 4-folds with singularity ranks lower than six, without a modification of the argument; thus, a 4D F-theory with one or more U(1) factors is obtained when our method is applied to construct elliptic Calabi--Yau 4-folds as double covers of 1/2 Calabi--Yau 4-folds with singularity ranks (strictly) lower than six. 
\par The double covers of 1/2 Calabi--Yau 4-folds constructed as ``duals'' of the cubic hypersurfaces with singularity types $D_5$, $A_4A_1$, and $A_4$ in $\P^3$ yield Calabi--Yau 4-folds with singularity types $D_5$, $A_4A_1$, and $A_4$. F-theory compactified on such Calabi--Yau 4-folds can yield theories whose gauge groups are relevant to a grand unified theory (GUT). However, it is necessary to determine whether the singular fibers corresponding to these singularity types are split/semi-split/non-split, to confirm whether an $SU(5)$ or $SO(10)$ gauge group is formed in 4D F-theory \cite{BIKMSV}. It may be interesting to investigate further details of these models in future studies.

\vspace{5mm}

\par The family of Calabi--Yau 4-folds constructed in \cite{Kimura1911} as the double covers of 1/2 Calabi--Yau 4-folds generate non-Abelian gauge group factors of ranks up to six. Are there Calabi--Yau 4-folds whose bases are isomorphic to Fano 3-folds of degree two, on which F-theory compactifications generate non-Abelian gauge group factors of ranks higher than six? The condition that the base 3-fold of an elliptic Calabi--Yau 4-fold is isomorphic to a Fano 3-fold of degree two does not seem to preclude this, based on reasoning that the condition imposed on the geometry of base 3-fold is not much strong; thus, it seems natural to expect that there are Calabi--Yau 4-folds on which F-theory compactification provides 4D theories with non-Abelian gauge groups of ranks greater than six. Do the geometric properties of the Calabi--Yau 4-folds that permit their being split into building blocks of rational elliptic 4-folds also impose constraints on the ranks of the (non-Abelian) gauge groups formed in F-theory? Studying this can assist in analyzing the structure of 4D $N=1$ F-theory moduli in relation to the swampland conditions.

\vspace{5mm}

\par The method that we employed in this work provided a means of deducing the configurations of the base points that can be blown up to yield 1/2 Calabi--Yau 4-folds, as well as the equations of the four (1,1) hypersurfaces that specify the base points. Our method can be used to classify the singularity types of the 1/2 Calabi--Yau 4-folds. Furthermore, the blow-up technique that we utilized in section \ref{sec2.2} can be used to analyze the non-Abelian gauge groups formed in 4D F-theory on the Calabi--Yau 4-folds constructed as double covers. This technique might also be useful in studying matter spectra at the intersections of the 7-branes, as well as Yukawa couplings. However, the presence of four-form flux influences the gauge groups and matter spectra in 4D F-theory \cite{MTsection}. A future study might focus upon the matter spectra generated in 4D F-theory when applied to the Calabi--Yau 4-folds constructed as the double covers of 1/2 Calabi--Yau 4-folds.

\section{Concluding remarks and open problems}
\label{sec4}
In this work, we introduced a method of classifying the singularity types of 1/2 Calabi--Yau 4-folds. This method also classifies the singularity types of the Calabi--Yau 4-folds constructed as the double covers of the 1/2 Calabi--Yau 4-folds. The types of non-Abelian gauge groups formed on the 7-branes in 4D F-theory on such Calabi--Yau 4-folds can be deduced from the singularity types. 
\par We explicitly analyzed $3A_2$ and $D_4$ singularity types as a demonstration of our method. The case of the $E_6$ singularity is complicated, and the question of whether 1/2 Calabi--Yau 4-folds with an $E_6$ singularity exist or not remains undetermined. 
\par Our method applies equally well to 1/2 Calabi--Yau 4-folds with singularity ranks other than six and four. Analyzing 1/2 Calabi--Yau 4-folds with singularity types besides $3A_2$ and $D_4$, as well as those with Calabi--Yau 4-folds as their double covers, represents a future research direction. 4D F-theory on the double covers of 1/2 Calabi--Yau 4-folds with singularity types $D_5$, $A_4A_1$, and $A_4$ may be relevant to GUT models. 
\par If the matrix representations of dual cubic hypersurfaces in $\P^3$ of 1/2 Calabi--Yau 4-folds are determined, then the equations of the four (1,1) hypersurfaces yielding 1/2 Calabi--Yau 4-folds can be deduced from these matrix representations. The blow-up operation reveals the structures of the singular fibers. Taking the double covers of the studied 1/2 Calabi--Yau 4-folds yields Calabi--Yau 4-folds, on which F-theory compactifications provide 4D $N=1$ theories. 
\par When one can construct a 1/2 Calabi--Yau 4-fold whose singularity rank is less than or equal to five, 4D F-theory construction on the Calabi--Yau double cover has at least one U(1) factor \cite{Kimura1911}. 
\par The blow-ups discussed in section \ref{sec2.2} describe sections that a 1/2 Calabi--Yau 4-fold possesses. The sections arising from blow-ups at the base points of four (1,1) hypersurfaces generate the Mordell--Weil group \cite{Kimura1910, Kimura1911}. Because a base change of the sections of the 1/2 Calabi--Yau 4-folds yields sections of the Calabi--Yau 4-fold double cover \cite{Kimura1911}, the method of blow-up discussed in section \ref{sec2.2} can be used to study the U(1) gauge group formed in 4D F-theory on a Calabi--Yau 4-fold constructed as the double cover. 
\par Studies of matter spectra and Yukawa couplings in 4D F-theory on the Calabi--Yau 4-fold double covers of 1/2 Calabi--Yau 4-folds are left for future studies. The Weierstrass equations of 1/2 Calabi--Yau 4-folds (and those of the Calabi--Yau 4-folds constructed as their double covers) assist in analyzing these; however, deducing the Weierstrass equation from the given equations of the four (1,1) hypersurfaces that yield a 1/2 Calabi--Yau 4-fold is in most cases considerably difficult. In analyzing gauge groups and matter fields arising in 4D F-theory, it is beneficial to find an algorithm to deduce the Weierstrass equation from the given equations of the four (1,1) hypersurfaces in $\P^2\times\P^2$.
\par The splitting of a Calabi--Yau 4-fold into a pair of 1/2 Calabi--Yau 4-folds can be viewed as a 4D analogue of the stable degeneration limit \cite{FMW, AM}. In the moduli of elliptically fibered Calabi--Yau 4-folds, do those that permit splitting into a pair of rational elliptic 4-folds correspond to some special limits of physical meaning? When the base degree-two Fano 3-fold of the Calabi--Yau 4-fold (constructed as the double cover of a 1/2 Calabi--Yau 4-fold) admits a conic fibration, it is natural to expect that the Calabi--Yau 4-fold also has a K3 fibration that is compatible with the elliptic fibration, as hypothesized in \cite{Kimura1911}. If this is true, then because 4D F-theory on a K3-fibered elliptic Calabi--Yau 4-fold has a heterotic dual \cite{Vaf, MV1, MV2, Sen, FMW}, the splitting limit can also be analyzed from the dual heterotic perspective.

\section*{Acknowledgments}

We would like to thank Shigeru Mukai for discussions.

\end{document}